\documentclass[%
 reprint,
superscriptaddress,
 amsmath,amssymb,
 aps, prl
]{revtex4-1}

\usepackage{graphicx}
\DeclareGraphicsExtensions{.pdf}
\graphicspath{ {images/} }

\usepackage{dcolumn}
\usepackage{bm}
\usepackage{natbib}
\usepackage{hyperref}
\hypersetup{
    colorlinks = true,
    citecolor = {cyan},
    urlcolor = {cyan},
    linkcolor = {cyan},
}
\usepackage[mathlines]{lineno}

\begin{document}


\title{On Nematicity and Charge Order in Superoxygenated La$_{2-x}$Sr$_x$CuO$_{4+y}$}

\author{Zhiwei Zhang}
 \email{zhiwei.zhang@uconn.edu}
\affiliation{Department of Physics, University of Connecticut, Storrs CT 06269, USA}
\author{R. Sutarto}
\affiliation{Canadian Light Source, Saskatoon, Saskatchewan S7N 2V3, Canada}
\author{F. He}
\affiliation{Canadian Light Source, Saskatoon, Saskatchewan S7N 2V3, Canada}
\author{F. C. Chou}
\affiliation{Center for Condensed Matter Sciences, National Taiwan University, Taipei 10670, Taiwan}
\author{L. Udby}
\affiliation{X-ray and Neutron Science, Niels Bohr Institute, University of Copenhagen, DK-2100 Copenhagen, Denmark}
\author{S. L. Holm}
\affiliation{X-ray and Neutron Science, Niels Bohr Institute, University of Copenhagen, DK-2100 Copenhagen, Denmark}
\author{Z. H. Zhu}
\affiliation{Department of Physics, Massachusetts Institute of Technology, Cambridge MA 02139, USA}
\author{W. A. Hines}
\affiliation{Department of Physics, University of Connecticut, Storrs CT 06269, USA}
\author{J. I. Budnick}
\affiliation{Department of Physics, University of Connecticut, Storrs CT 06269, USA}
\author{B. O. Wells}
\affiliation{Department of Physics, University of Connecticut, Storrs CT 06269, USA}

\date{\today}


\begin{abstract}

 A nematic order in stripe-ordered cuprates was recently identified with (001) reflection at resonant energies associated with the in-plane states. However, whether this resonant reflection is ubiquitous among all 214 cuprates is still unknown. Here we report a Resonant soft X-ray Scattering (RXS) measurement on two La$_{2-x}$Sr$_x$CuO$_{4+y}$ crystals. Charge order was found in La$_2$CuO$_{4+y}$ with a total hole concentration near 0.125/Cu but no measurable (001) peak at any resonance, while in a La$_{1.94}$Sr$_{0.06}$CuO$_{4+y}$ sample near 0.16/Cu a (001) peak resonant was identified to be consistent with the presence of LTT tilting. The lack of such a (001) peak in a compound with stripe-like charge order raises questions about nematicity and the origin of the scattering feature.
 
\end{abstract}

\maketitle

An ongoing, critical issue concerning cuprate superconductors is the extent to which charge stripe order is identified with an electronic nematic state: an orientational ordering of the conduction electrons that breaks the symmetry of the lattice.\cite{Kivelson1998} While the presence of electronic nematic order is now well accepted in the Fe-compound superconductors,\cite{Chu824} in cuprates such order is expected to alternate directions plane by plane making it difficult to measure by transport. A recent manuscript reports a clean measure of nematic order in 214 cuprates using resonant scattering: the detection of the nominally disallowed (001) peak under resonance at energies associated with in-plane Cu-O states.\cite{Achkar576} Whether this interpretation of the resonant (001) holds generally is not yet known.

The advent of powerful resonant scattering techniques has allowed for the detection of charge order (CO).\cite{PhysRevB.90.054513, PhysRevB.79.100502, daSilvaNeto24012014, Comin390, daSilvaNeto282, Wu2012, Abbamonte2005} In 214
cuprates, an interwoven concomitant charge and spin stripe-like order has been known for over two decades,\cite{PhysRevB.40.7391, PhysRevB.39.9749, MACHIDA1989192}
but was originally only reported in samples with the low temperature tetragonal (LTT, $P4_2/ncm$)\cite{Tranquada1995, PhysRevB.79.100502} 
or low temperature less orthorhombic structure (LTLO)\cite{PhysRevB.91.054521}. More recently, CO has been found in nearly all cuprate families doped near 1/8th, but in non-214 compounds the CO has not matched a stripe model of charges and spins.

\begin{figure*}
    \centering
    \includegraphics[width=0.98\linewidth]{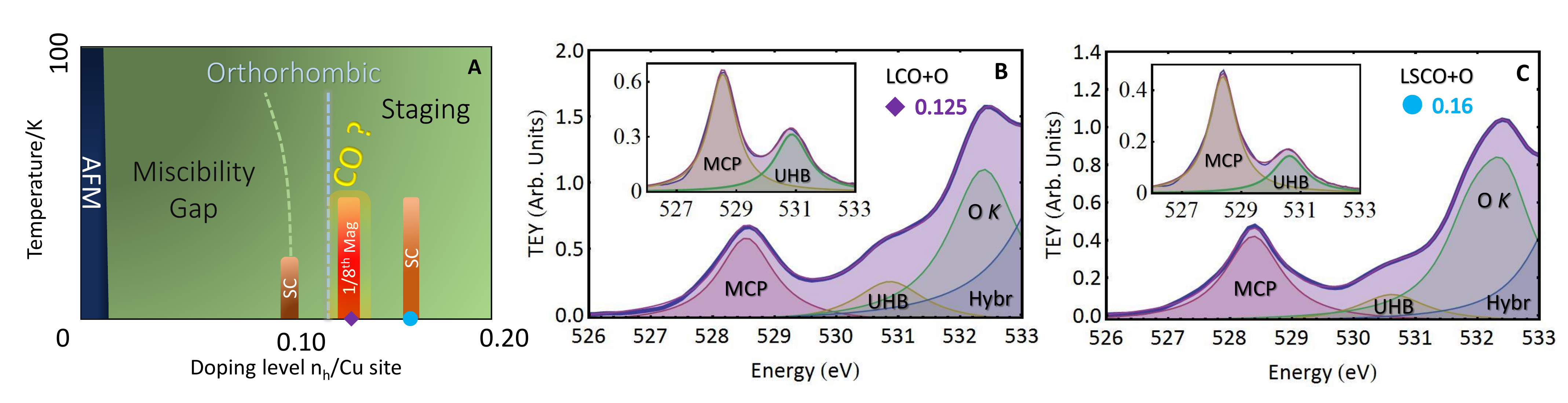}
    \caption{\textbf{A}. Phase diagram for superoxygenated lanthanum cuprates. Though 1/8th magnetic correlation exists, the existence of CO remains unclear for many years. The doping levels for the samples studied in this paper are indicated by the diamonds in the phase diagram. \textbf{B(C)}. Fitting profile for the total electron yield of the O-XAS for LCO+O(LSCO+O). The insets are the pre-edge peaks MCP and UHB for the two samples obtained by subtracting the $K$ edges and hybridization from the raw data, respectively. Corresponding symbols are used to indicate the doping levels for both samples in the phase diagram in \textbf{A}.}
    \label{1}
\end{figure*}

A unique material system that should be particularly well suited to studying charge and spin order associated with the 1/8th doped phases is superoxygenated La$_{2-x}$Sr$_x$CuO$_{4+y}$, i.e. La$_2$CuO$_4$ co-doped with Sr on La sites and interstitial oxygen. As shown in Fig.~\ref{1}A, this system exhibits inherent electronic phase separation, with large regions of the sample favoring the 1/8th doped magnetically striped state and other regions the optimally doped superconductor. Given the current understanding of CO, the self-segregated 1/8th doped phase should be a clean example of the CO material. Indeed, neutron\cite{PhysRevLett.111.227001} and $\mu$SR\cite{Mohottala2006} studies have shown that the magnetic state is very well ordered despite the fact that the compound remains in the low temperature orthorhombic phase (LTO, $Bmab$), which does not have any structural elements that would obviously favor stripes.\cite{PhysRevLett.111.227001} Here we report the discovery of charge order using RXS in a sample of LCO+O with a total charge doping level near 1/8th. We find resonant scattering peaks with the periodicity, temperature and resonant behavior expected for stripe-like CO. However, the (001) resonant peak associated with nematic order is not present in this same sample. This raises the prospect of having stripe-like CO without nematic orientational order, a combination that is difficult to reconcile with current theories.

This Letter focuses on the data from two crystals. Charge order peaks appear in a flux grown La$_2$CuO$_4$ sample oxidized for over 80 days 
(LCO+O) using a wet electrochemical method. The other crystal was grown
in a light furnace using the travelling solvent floating zone technique and oxidized for approximately a year. This sample (LSCO+O) was co-doped with Sr and O, with 6\% Sr
on La sites. Both LCO+O and LSCO+O were superconductors with a sharp transition $T_\text{c}$ of 40 K. The LCO+O was cleaved in air then immediately transferred to the vacuum and cooled to 20 K whereas the LSCO+O was cleaved at low temperature in vacuum.

A critical issue for the presence of charge and spin order is the local hole density, determined by Sr and O doping. Knowing the detailed oxygen concentration is difficult without performing destructive testing such as thermal gravimetric analysis.\cite{Mohottala2006} More importantly, for a near-surface techniques such as RXS, the region sampled may not have the same oxygen concentration as the bulk.
The best measure of the doping level in the region of interest is the O $K$-edge absorption spectrum measured at the same time as the scattering. We use the total electron yield (TEY) measure, as it is less subject to saturation effects than fluorescence measurements and probes a near-surface region (tens of nanometers) safely within the range probed by resonant scattering. There are two pre-peaks to the main O $K$ edge, the first identified as the mobile carrier peak (MCP) and the second as the upper Hubbard band (UHB). It has been shown that with hole doping, the MCP grows in intensity while the UHB reduces. The ratio of the two is a measure of the hole concentration in the region sampled, with some complication in carrying out an accurate subtraction of the contribution from the rest of the K edge. Fig.~\ref{1}B(C) shows the oxygen $K$-edge x-ray absorption for both samples measured in the TEY. The MCP and UHB peaks are at $\sim$528 and $\sim$531 eV, followed by the main part of the $K$ edge above 532 eV. By comparing the shapes of the whole spectra to that presented by Chen \textit{et al.}\cite{ctchen}, we found that the doping levels for our LCO+O and LSCO+O samples are $\sim 0.125$ and $\sim 0.16$, respectively. More details can be found in the Supplemental Material\cite{supp}. 

   \begin{figure}[h]
    \centering
    \includegraphics[width=0.98\linewidth]{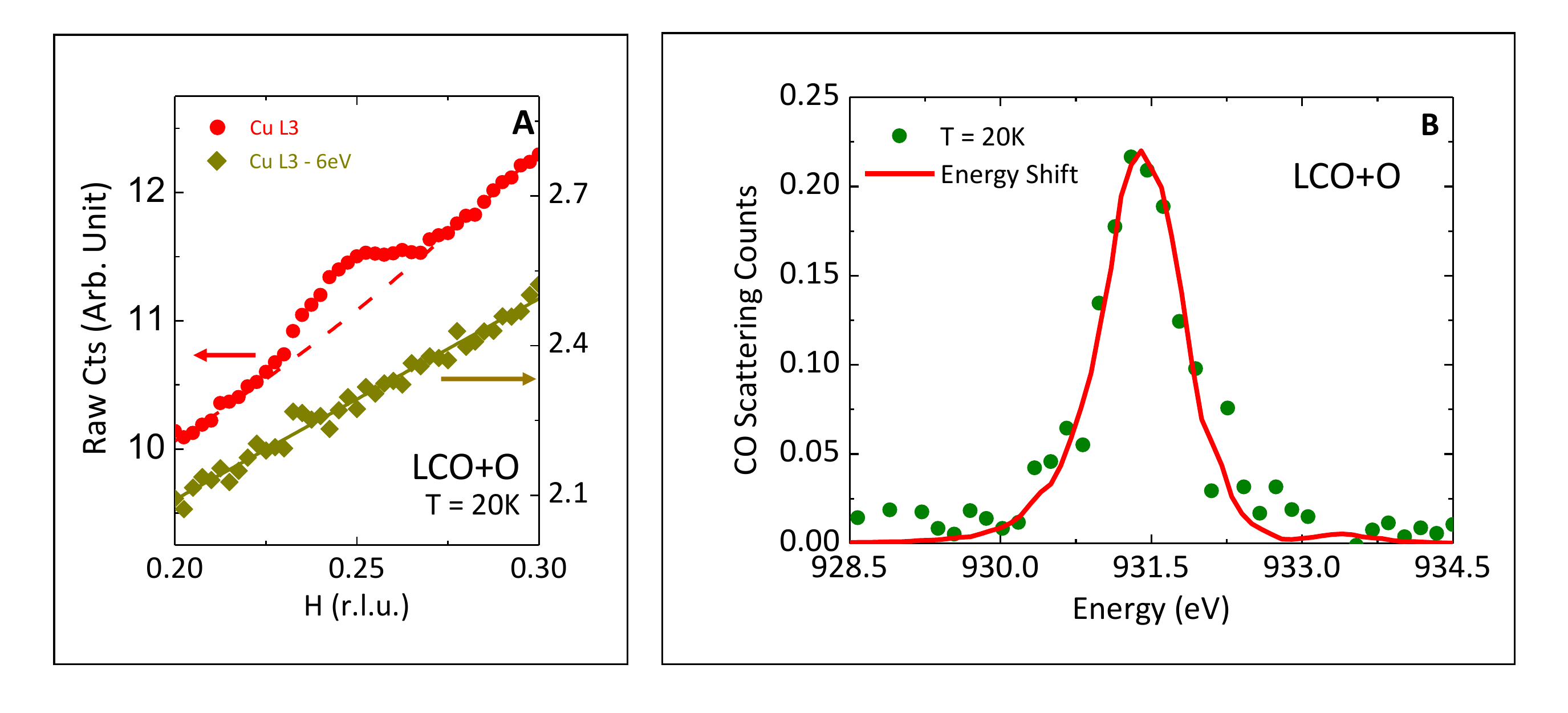}
    \caption{\textbf{A}. The appearance of CO peak on resonance and the non-appearance off resonance. The red dashed is the fitted background. \textbf{B}. Energy dependence of the CO peak intensity at 20 K and calculated energy-dependence by the energy-shift model.}
    \label{2}
   \end{figure}
 
 Fig.~\ref{2}A shows the region where we expect to find a CO peak on and off resonance at the Cu $L_3$ edge at $T=20$K. A peak appears on resonance only. The energy profile of the scattering peak closely matches that of the Cu $L_3$ XAS itself. The red curve is a fit to the data using the energy shift model\cite{PhysRevLett.110.017001}. This model postulates that the absorption for the on-stripe and off-stripe Cu atoms differ only by a small energy shift. In both cases, the model fits the CO scattering peak intensity across the Cu $L_3$ edge well. The Cu XAS at 60 K is shown in Fig. S3A and the form factors used in the fits are given in Fig. S3B.\citep{supp} At the O $K$ edge, we find only a hint of a CO peak that cannot clearly be detected above the background. The CO resonance on oxygen edges (particularly MCP) is either weak or absent.
   
\begin{figure}[h]
    \centering
    \includegraphics[width=0.98\linewidth]{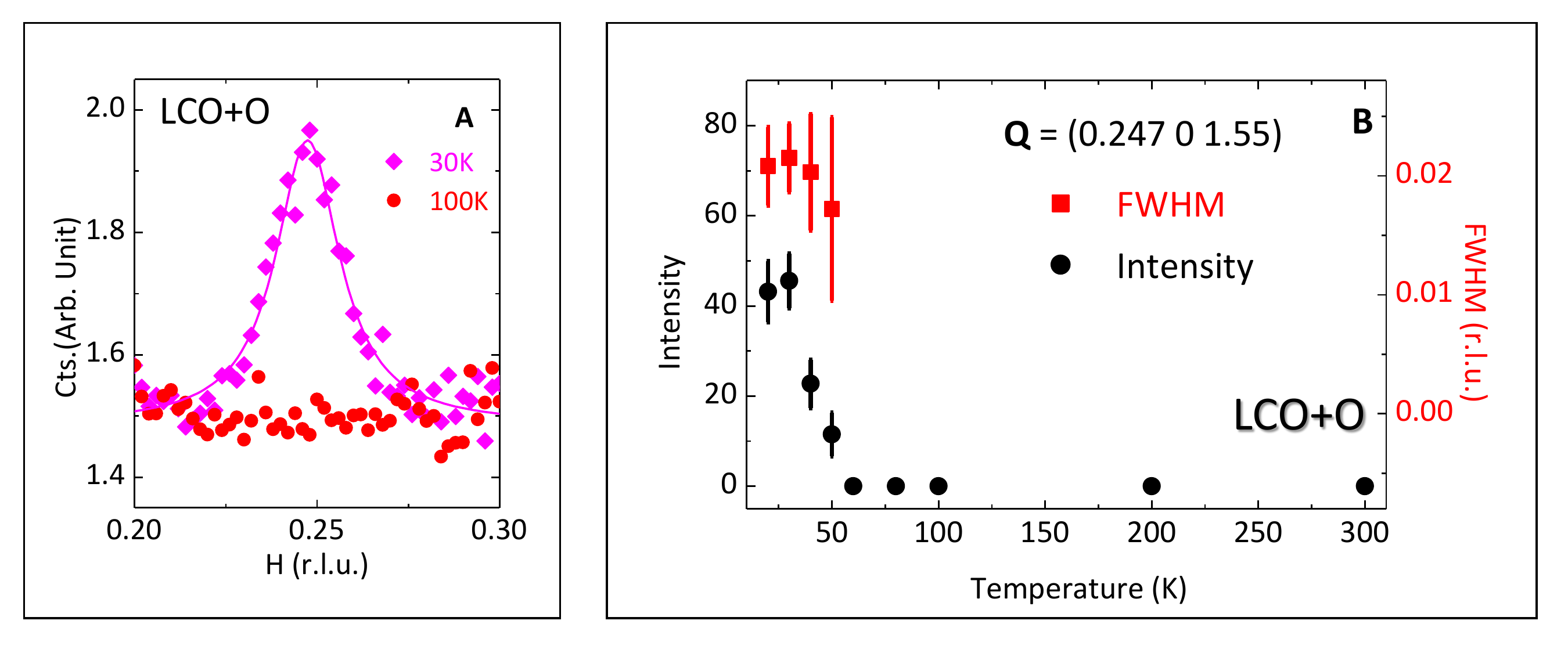}
    \caption{\textbf{A}. The appearance of the resonant CO peak of ($H$ 0 1.55) at 30 K and the disappearance at 100 K. Background was subtracted. \textbf{B}. The temperature dependence of the integrated intensity and the width in $H$ for the CO peak in LCO+O on Cu $L_3$ edge.}
    \label{3}
  \end{figure}

 In Fig.~\ref{3} we show the temperature dependence of the CO peak. Fig.~\ref{3}A shows the background-subtracted CO peak at 30 K and 100 K while
 Fig.~\ref{3}B shows the full temperature dependence of the peak intensity and width taken from fitting the peaks. It is clear that the
 transition temperature of CO is $\sim 50$K. A complete set of scans at different temperatures are presented in the Supplemental Material\cite{supp}. The peak width remains constant below the transition temperature while the intensity grows like an order parameter as the sample 
 is cooled. At 20 K the peak width corresponds to a correlation length for CO of 60$\text{\AA}$ ($=1/\text{HWHM}$), 5 times shorter than magnetic
 correlation length reported in Ref.\cite{PhysRevLett.111.227001}. Different than in most cuprates, the intensity of the CO peak does not
 drop at the superconducting transition temperature and thus in this manner we don't see competition between CO and superconductivity.
 However, in the superoxygenated compounds, the superconducting and magnetic regions phase separate and exist in 
 different regions of the sample, thus already satisfying the competition between charge/spin order and superconductivity. 
 
The parameters of the CO in the charge ordered sample appear
roughly as one would expect assuming both the ubiquity of stripe-like charge order in $n_h =$ 1/8th 214 cuprates and the presence
of a substantial fraction of the LCO+O sample separated into the 1/8th doped phase. The low temperature correlation length of 60{\AA} is
smaller than in similar samples that showed similar resolution limited magnetic neutron scattering peaks. The ordering temperatures for both charge and spin order are remarkably similar to that found in 1/8th doped LBCO, 40 K for spin order and 55 K for charge order. The latter is surprising since the charge order transition coincides with the LTO to LTT structural transition in LBCO\cite{PhysRevB.83.104506} but there is no such transition in the superoxygenated LCO+O. The similarity might imply that the charge ordering in LBCO is not simply determined by the structural transition and the transition temperature may be more intrinsic.

\begin{figure*}
    \centering
    \includegraphics[width=0.99\linewidth]{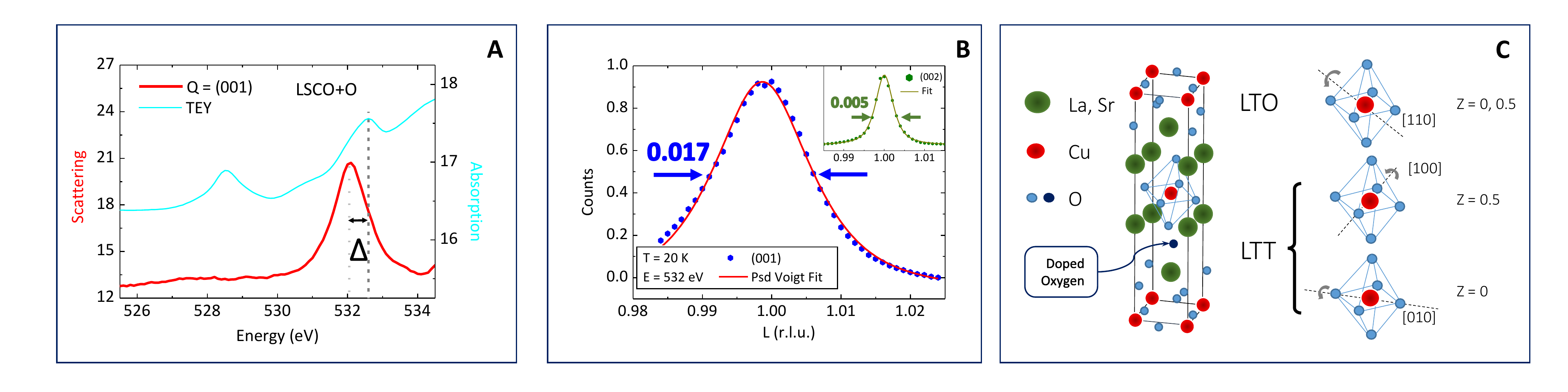}
    \caption{\textbf{A}. Energy dependence of the (001) reflection in LSCO+O  at 20 K. The data plotted is a constant $\bm{Q} =$ (001) energy scan with a background subtracted from a subsequent scan with the detector out of the scattering plane. The scattering response and the TEY are in red and cyan colors, respectively. The scales on the two vertical axes are not comparable. \textbf{B}. $L$-scan of the (001) at the most resonance (532 eV). The higher order (002) contamination was subtracted from a measurement off resonance. \textbf{C}. CuO$_6$ tilt patterns for both LTO and LTT structures. The LTT-type tilting causes anisotropy between neighboring layers while the LTO-type does not.}
    \label{4}
 \end{figure*}

Previous work has found that the normally disallowed (001) peak is present on resonance in 1/8th doped, 214 charge ordered samples. Some time ago it was shown that in the presence of the type of CuO$_6$ octahedral tilts that characterize the LTT phase, an electronic ordering in the hybridized states between apical O and La makes this peak appear on resonance.\cite{PhysRevB.83.092503} More recently, an additional resonant energy profile was found for the (001) peak that had a temperature profile associated with charge ordering.\cite{Achkar576} A possible complication in measuring the (001) peak especially near O $K$ edges is higher order light leading to the (002) reflection at the same spectrometer position. However, a constant-$\bm{Q}$ energy scan allows us to separate the two contributions as there is no possible resonance of the (002) peak at $\lambda$/2 in the region where $\lambda$ is near the O $K$ edges. We find that our sample with charge order has no measurable resonant (001) peak on Cu $L$, La $M$ or O $K$ edges. However, the LSCO+O sample with higher hole concentration and no charge order does have a resonant (001) reflection near O $K$ edge which is robust up to at least 70 K, well above the transition temperature for CO in LCO+O. In Fig.~\ref{4}A, the energy dependence of the (001) reflection is plotted in the region of the O $K$ edge. The data plotted is a constant $\bm{Q} =$ (001) energy scan with a background subtracted from a subsequent scan with the detector out of the scattering plane. This leaves intensity from both the (001) peak and the (002) with higher order light, but only the former will have a resonant profile. For reference, the XAS profile measured with TEY is also plotted in the figure. The resonant (001) peak profile in $Q$ can also be extracted, which is shown in Fig.~\ref{4}B. The (001) peak is about three times broader than the (002) peak (shown in the inset of Fig.~\ref{4}B), indicating that the resonant (001) peak represents an ordering that extends over a significantly smaller region than the crystalline order itself. Technical details concerning the extraction of the (001) peak are in the Supplemental Material\cite{supp}.

The behavior of the (001) peak in both samples is surprising.  The lack of an (001) peak resonant at the Cu and in-plane oxygen state energies differs from the result reported by Achkar \textit{et al}. in similar compounds. Those authors interpret this resonant peak as arising from an electronic nematic associated with stripe-like charge and spin order. This conclusion is drawn from their calculation of the peak structure factor which gives an  intensity on resonance proportional to $\eta^2$, with $\eta = f_{aa}(z=0) - f_{aa}(z=0.5)$. Invoking the symmetry of the structure gives $f_{aa}(z=0.5)=f_{bb}(z=0)$. Combining the two results in $\eta = f_{aa}(z=0) - f_{bb}(z=0)$, and thus the peak intensity is given by differences between the electronic states in the two principle in-plane directions. In our case, the lack of such a peak must mean one of the following possibilities holds.

One possibility is that in our charge-ordered sample $f_{aa}(z=0)=f_{bb}(z=0)$, and thus $\eta=0$. This condition is incompatible with charge stripes as usually conceived, but would allow for the sort of checkerboard pattern that Christensen \textit{et al.} found to be a compatible spin structure for their set of neutron magnetic peaks.\cite{PhysRevLett.98.197003}

The other is that the symmetry condition $f_{aa}(z=0.5)=f_{bb}(z=0)$ does not hold in our sample. This would imply the presence of spin stripes that do not alternate direction ($\hat{\bm{a}}$ vs. $\hat{\bm{b}}$) in adjacent CuO$_2$ layers. 
Neutron scattering study of the spin order in a set of superoxygenated samples found equal intensities for all four of the set of incommensurate spin order peaks, representing equal populations of stripes along $\hat{\bm{a}}$ and along $\hat{\bm{b}}$.\cite{PhysRevLett.111.227001} That suggests some ordering of stripe orientation to enforce equal populations. In addition, in this work the charge order peak is found to be most prominent near $L=$1.5. Half integer values for CO peaks implies a next-nearest neighbor of the charge order which in other 214 cuprates is built upon the alternating stripe direction.\cite{zimmerman}

While there is no detectable (001) peak in the LCO+O sample with charge order, we do detect the (001) peak on resonance in the more heavily
doped LSCO+O sample. This observation gives confidence that the null result in the charge order sample is robust, but in itself is a surprising
observation. The energy dependence for the (001) peak is very much like that previously published in LESCO in the overdoped region.\cite{PhysRevB.83.092503} 
Ray \textit{et al.} found that the 6.5\% Sr superoxygenated samples have staging peaks with significantly broader tails than the 4\% samples.\cite{1707.08871} Staging is related to the octahedral tilts with some ordering patterns in superoxyenated samples.\cite{Wells1996} The broadened tails may be an indication of tilts around an axis other than the orthorhombic (010), a partial LTT or LTLO ordering. In LBCO, the presence of the LTT plus LTLO phase is considered to pin the charge order, and static stripe-like spin and charge order is measured up to the optimal doping region. It appears that the degree of LTT/LTLO indicated by the (001) resonance in our sample is not enough to pin charge order which appears confined to the phase separated 1/8th doped regions.

In conclusion, we report RXS study on two superoxygenated 214 cuprates, one doped only with oxygen leading to a hole concentration near 1/8th and the other co-doped with Sr and oxygen with a hole concentration near 0.16. While both are phase separated into magnetic and superconducting ($T_\text{c}$=40 K) regions, the former should be primarily magnetic and the latter primarily superconducting. Charge order, found in the 1/8th doped sample, has a transition temperature of 50 K, similar to several other 214 cuprates. We note that spins in these samples order near 40 K, which appears to be near-universal.\cite{PhysRevLett.111.227001, PhysRevB.57.6165, PhysRevB.83.104506, PhysRevB.78.174529, PhysRevLett.85.4590} In most 214 compounds, charge order were found in LTT or LTLO phase, or even at twin domain boundaries of LTT-like tilts in La$_{15/8}$Sr$_{1/8}$CuO$_4$ compounds\citep{Wu2012,1404.3192,PhysRevB.89.224513}. In our LCO+O sample, while we cannot rule out that the charge order is confined to the near surface or domain walls, we can rule out LTT-like tilts by the absence of a resonant (001) peak at the apical oxygen or La edges. Thus while stripe-like charge and spin order remains particular to the 214-type cuprates, it may not be closely tied to symmetry breaking structural tilt patterns as previously believed.

We acknowledge helpful conversations with Ch. Niedermayer and M. H\"ucker. We thank Y. Zhou, D. Paley and Y. Lin for their help with sample crystal orientations. Research described in this paper was performed at the Resonant Elastic and Inelastic X-ray Scattering (REIXS) Beamline of the Canadian Light Source, which is supported by the Canada Foundation for Innovation, Natural Sciences and Engineering Research Council of Canada, the University of Saskatchewan, the Government of Saskatchewan, Western Economic Diversification Canada, the National Research Council Canada, and the Canadian Institutes of Health Research. Work at the University of Connecticut was supported by DOE-BES Contract No. DEFG02-00ER45801.

\bibliographystyle{apsrev4-1}

\begin{thebibliography}{31}%
\makeatletter
\providecommand \@ifxundefined [1]{%
 \@ifx{#1\undefined}
}%
\providecommand \@ifnum [1]{%
 \ifnum #1\expandafter \@firstoftwo
 \else \expandafter \@secondoftwo
 \fi
}%
\providecommand \@ifx [1]{%
 \ifx #1\expandafter \@firstoftwo
 \else \expandafter \@secondoftwo
 \fi
}%
\providecommand \natexlab [1]{#1}%
\providecommand \enquote  [1]{``#1''}%
\providecommand \bibnamefont  [1]{#1}%
\providecommand \bibfnamefont [1]{#1}%
\providecommand \citenamefont [1]{#1}%
\providecommand \href@noop [0]{\@secondoftwo}%
\providecommand \href [0]{\begingroup \@sanitize@url \@href}%
\providecommand \@href[1]{\@@startlink{#1}\@@href}%
\providecommand \@@href[1]{\endgroup#1\@@endlink}%
\providecommand \@sanitize@url [0]{\catcode `\\12\catcode `\$12\catcode
  `\&12\catcode `\#12\catcode `\^12\catcode `\_12\catcode `\%12\relax}%
\providecommand \@@startlink[1]{}%
\providecommand \@@endlink[0]{}%
\providecommand \url  [0]{\begingroup\@sanitize@url \@url }%
\providecommand \@url [1]{\endgroup\@href {#1}{\urlprefix }}%
\providecommand \urlprefix  [0]{URL }%
\providecommand \Eprint [0]{\href }%
\providecommand \doibase [0]{http://dx.doi.org/}%
\providecommand \selectlanguage [0]{\@gobble}%
\providecommand \bibinfo  [0]{\@secondoftwo}%
\providecommand \bibfield  [0]{\@secondoftwo}%
\providecommand \translation [1]{[#1]}%
\providecommand \BibitemOpen [0]{}%
\providecommand \bibitemStop [0]{}%
\providecommand \bibitemNoStop [0]{.\EOS\space}%
\providecommand \EOS [0]{\spacefactor3000\relax}%
\providecommand \BibitemShut  [1]{\csname bibitem#1\endcsname}%
\let\auto@bib@innerbib\@empty
\bibitem [{\citenamefont {Kivelson}\ \emph {et~al.}(1998)\citenamefont
  {Kivelson}, \citenamefont {Fradkin},\ and\ \citenamefont
  {Emery}}]{Kivelson1998}%
  \BibitemOpen
  \bibfield  {author} {\bibinfo {author} {\bibfnamefont {S.~A.}\ \bibnamefont
  {Kivelson}}, \bibinfo {author} {\bibfnamefont {E.}~\bibnamefont {Fradkin}}, \
  and\ \bibinfo {author} {\bibfnamefont {V.~J.}\ \bibnamefont {Emery}},\ }\href
  {\doibase 10.1038/31177} {\bibfield  {journal} {\bibinfo  {journal} {Nature}\
  }\textbf {\bibinfo {volume} {393}},\ \bibinfo {pages} {550} (\bibinfo {year}
  {1998})}\BibitemShut {NoStop}%
\bibitem [{\citenamefont {Chu}\ \emph {et~al.}(2010)\citenamefont {Chu},
  \citenamefont {Analytis}, \citenamefont {De~Greve}, \citenamefont {McMahon},
  \citenamefont {Islam}, \citenamefont {Yamamoto},\ and\ \citenamefont
  {Fisher}}]{Chu824}%
  \BibitemOpen
  \bibfield  {author} {\bibinfo {author} {\bibfnamefont {J.-H.}\ \bibnamefont
  {Chu}}, \bibinfo {author} {\bibfnamefont {J.~G.}\ \bibnamefont {Analytis}},
  \bibinfo {author} {\bibfnamefont {K.}~\bibnamefont {De~Greve}}, \bibinfo
  {author} {\bibfnamefont {P.~L.}\ \bibnamefont {McMahon}}, \bibinfo {author}
  {\bibfnamefont {Z.}~\bibnamefont {Islam}}, \bibinfo {author} {\bibfnamefont
  {Y.}~\bibnamefont {Yamamoto}}, \ and\ \bibinfo {author} {\bibfnamefont
  {I.~R.}\ \bibnamefont {Fisher}},\ }\href {\doibase 10.1126/science.1190482}
  {\bibfield  {journal} {\bibinfo  {journal} {Science}\ }\textbf {\bibinfo
  {volume} {329}},\ \bibinfo {pages} {824} (\bibinfo {year}
  {2010})}\BibitemShut {NoStop}%
\bibitem [{\citenamefont {Achkar}\ \emph {et~al.}(2016)\citenamefont {Achkar},
  \citenamefont {Zwiebler}, \citenamefont {McMahon}, \citenamefont {He},
  \citenamefont {Sutarto}, \citenamefont {Djianto}, \citenamefont {Hao},
  \citenamefont {Gingras}, \citenamefont {H{\"u}cker}, \citenamefont {Gu},
  \citenamefont {Revcolevschi}, \citenamefont {Zhang}, \citenamefont {Kim},
  \citenamefont {Geck},\ and\ \citenamefont {Hawthorn}}]{Achkar576}%
  \BibitemOpen
  \bibfield  {author} {\bibinfo {author} {\bibfnamefont {A.~J.}\ \bibnamefont
  {Achkar}}, \bibinfo {author} {\bibfnamefont {M.}~\bibnamefont {Zwiebler}},
  \bibinfo {author} {\bibfnamefont {C.}~\bibnamefont {McMahon}}, \bibinfo
  {author} {\bibfnamefont {F.}~\bibnamefont {He}}, \bibinfo {author}
  {\bibfnamefont {R.}~\bibnamefont {Sutarto}}, \bibinfo {author} {\bibfnamefont
  {I.}~\bibnamefont {Djianto}}, \bibinfo {author} {\bibfnamefont
  {Z.}~\bibnamefont {Hao}}, \bibinfo {author} {\bibfnamefont {M.~J.~P.}\
  \bibnamefont {Gingras}}, \bibinfo {author} {\bibfnamefont {M.}~\bibnamefont
  {H{\"u}cker}}, \bibinfo {author} {\bibfnamefont {G.~D.}\ \bibnamefont {Gu}},
  \bibinfo {author} {\bibfnamefont {A.}~\bibnamefont {Revcolevschi}}, \bibinfo
  {author} {\bibfnamefont {H.}~\bibnamefont {Zhang}}, \bibinfo {author}
  {\bibfnamefont {Y.-J.}\ \bibnamefont {Kim}}, \bibinfo {author} {\bibfnamefont
  {J.}~\bibnamefont {Geck}}, \ and\ \bibinfo {author} {\bibfnamefont {D.~G.}\
  \bibnamefont {Hawthorn}},\ }\href {\doibase 10.1126/science.aad1824}
  {\bibfield  {journal} {\bibinfo  {journal} {Science}\ }\textbf {\bibinfo
  {volume} {351}},\ \bibinfo {pages} {576} (\bibinfo {year}
  {2016})}\BibitemShut {NoStop}%
\bibitem [{\citenamefont {Blanco-Canosa}\ \emph {et~al.}(2014)\citenamefont
  {Blanco-Canosa}, \citenamefont {Frano}, \citenamefont {Schierle},
  \citenamefont {Porras}, \citenamefont {Loew}, \citenamefont {Minola},
  \citenamefont {Bluschke}, \citenamefont {Weschke}, \citenamefont {Keimer},\
  and\ \citenamefont {Le~Tacon}}]{PhysRevB.90.054513}%
  \BibitemOpen
  \bibfield  {author} {\bibinfo {author} {\bibfnamefont {S.}~\bibnamefont
  {Blanco-Canosa}}, \bibinfo {author} {\bibfnamefont {A.}~\bibnamefont
  {Frano}}, \bibinfo {author} {\bibfnamefont {E.}~\bibnamefont {Schierle}},
  \bibinfo {author} {\bibfnamefont {J.}~\bibnamefont {Porras}}, \bibinfo
  {author} {\bibfnamefont {T.}~\bibnamefont {Loew}}, \bibinfo {author}
  {\bibfnamefont {M.}~\bibnamefont {Minola}}, \bibinfo {author} {\bibfnamefont
  {M.}~\bibnamefont {Bluschke}}, \bibinfo {author} {\bibfnamefont
  {E.}~\bibnamefont {Weschke}}, \bibinfo {author} {\bibfnamefont
  {B.}~\bibnamefont {Keimer}}, \ and\ \bibinfo {author} {\bibfnamefont
  {M.}~\bibnamefont {Le~Tacon}},\ }\href {\doibase 10.1103/PhysRevB.90.054513}
  {\bibfield  {journal} {\bibinfo  {journal} {Phys. Rev. B}\ }\textbf {\bibinfo
  {volume} {90}},\ \bibinfo {pages} {054513} (\bibinfo {year}
  {2014})}\BibitemShut {NoStop}%
\bibitem [{\citenamefont {Fink}\ \emph {et~al.}(2009)\citenamefont {Fink},
  \citenamefont {Schierle}, \citenamefont {Weschke}, \citenamefont {Geck},
  \citenamefont {Hawthorn}, \citenamefont {Soltwisch}, \citenamefont {Wadati},
  \citenamefont {Wu}, \citenamefont {D\"urr}, \citenamefont {Wizent},
  \citenamefont {B\"uchner},\ and\ \citenamefont
  {Sawatzky}}]{PhysRevB.79.100502}%
  \BibitemOpen
  \bibfield  {author} {\bibinfo {author} {\bibfnamefont {J.}~\bibnamefont
  {Fink}}, \bibinfo {author} {\bibfnamefont {E.}~\bibnamefont {Schierle}},
  \bibinfo {author} {\bibfnamefont {E.}~\bibnamefont {Weschke}}, \bibinfo
  {author} {\bibfnamefont {J.}~\bibnamefont {Geck}}, \bibinfo {author}
  {\bibfnamefont {D.}~\bibnamefont {Hawthorn}}, \bibinfo {author}
  {\bibfnamefont {V.}~\bibnamefont {Soltwisch}}, \bibinfo {author}
  {\bibfnamefont {H.}~\bibnamefont {Wadati}}, \bibinfo {author} {\bibfnamefont
  {H.-H.}\ \bibnamefont {Wu}}, \bibinfo {author} {\bibfnamefont {H.~A.}\
  \bibnamefont {D\"urr}}, \bibinfo {author} {\bibfnamefont {N.}~\bibnamefont
  {Wizent}}, \bibinfo {author} {\bibfnamefont {B.}~\bibnamefont {B\"uchner}}, \
  and\ \bibinfo {author} {\bibfnamefont {G.~A.}\ \bibnamefont {Sawatzky}},\
  }\href {\doibase 10.1103/PhysRevB.79.100502} {\bibfield  {journal} {\bibinfo
  {journal} {Phys. Rev. B}\ }\textbf {\bibinfo {volume} {79}},\ \bibinfo
  {pages} {100502} (\bibinfo {year} {2009})}\BibitemShut {NoStop}%
\bibitem [{\citenamefont {da~Silva~Neto}\ \emph {et~al.}(2014)\citenamefont
  {da~Silva~Neto}, \citenamefont {Aynajian}, \citenamefont {Frano},
  \citenamefont {Comin}, \citenamefont {Schierle}, \citenamefont {Weschke},
  \citenamefont {Gyenis}, \citenamefont {Wen}, \citenamefont {Schneeloch},
  \citenamefont {Xu}, \citenamefont {Ono}, \citenamefont {Gu}, \citenamefont
  {Le~Tacon},\ and\ \citenamefont {Yazdani}}]{daSilvaNeto24012014}%
  \BibitemOpen
  \bibfield  {author} {\bibinfo {author} {\bibfnamefont {E.~H.}\ \bibnamefont
  {da~Silva~Neto}}, \bibinfo {author} {\bibfnamefont {P.}~\bibnamefont
  {Aynajian}}, \bibinfo {author} {\bibfnamefont {A.}~\bibnamefont {Frano}},
  \bibinfo {author} {\bibfnamefont {R.}~\bibnamefont {Comin}}, \bibinfo
  {author} {\bibfnamefont {E.}~\bibnamefont {Schierle}}, \bibinfo {author}
  {\bibfnamefont {E.}~\bibnamefont {Weschke}}, \bibinfo {author} {\bibfnamefont
  {A.}~\bibnamefont {Gyenis}}, \bibinfo {author} {\bibfnamefont
  {J.}~\bibnamefont {Wen}}, \bibinfo {author} {\bibfnamefont {J.}~\bibnamefont
  {Schneeloch}}, \bibinfo {author} {\bibfnamefont {Z.}~\bibnamefont {Xu}},
  \bibinfo {author} {\bibfnamefont {S.}~\bibnamefont {Ono}}, \bibinfo {author}
  {\bibfnamefont {G.}~\bibnamefont {Gu}}, \bibinfo {author} {\bibfnamefont
  {M.}~\bibnamefont {Le~Tacon}}, \ and\ \bibinfo {author} {\bibfnamefont
  {A.}~\bibnamefont {Yazdani}},\ }\href {\doibase 10.1126/science.1243479}
  {\bibfield  {journal} {\bibinfo  {journal} {Science}\ }\textbf {\bibinfo
  {volume} {343}},\ \bibinfo {pages} {393} (\bibinfo {year}
  {2014})}\BibitemShut {NoStop}%
\bibitem [{\citenamefont {Comin}\ \emph {et~al.}(2014)\citenamefont {Comin},
  \citenamefont {Frano}, \citenamefont {Yee}, \citenamefont {Yoshida},
  \citenamefont {Eisaki}, \citenamefont {Schierle}, \citenamefont {Weschke},
  \citenamefont {Sutarto}, \citenamefont {He}, \citenamefont {Soumyanarayanan},
  \citenamefont {He}, \citenamefont {Le~Tacon}, \citenamefont {Elfimov},
  \citenamefont {Hoffman}, \citenamefont {Sawatzky}, \citenamefont {Keimer},\
  and\ \citenamefont {Damascelli}}]{Comin390}%
  \BibitemOpen
  \bibfield  {author} {\bibinfo {author} {\bibfnamefont {R.}~\bibnamefont
  {Comin}}, \bibinfo {author} {\bibfnamefont {A.}~\bibnamefont {Frano}},
  \bibinfo {author} {\bibfnamefont {M.~M.}\ \bibnamefont {Yee}}, \bibinfo
  {author} {\bibfnamefont {Y.}~\bibnamefont {Yoshida}}, \bibinfo {author}
  {\bibfnamefont {H.}~\bibnamefont {Eisaki}}, \bibinfo {author} {\bibfnamefont
  {E.}~\bibnamefont {Schierle}}, \bibinfo {author} {\bibfnamefont
  {E.}~\bibnamefont {Weschke}}, \bibinfo {author} {\bibfnamefont
  {R.}~\bibnamefont {Sutarto}}, \bibinfo {author} {\bibfnamefont
  {F.}~\bibnamefont {He}}, \bibinfo {author} {\bibfnamefont {A.}~\bibnamefont
  {Soumyanarayanan}}, \bibinfo {author} {\bibfnamefont {Y.}~\bibnamefont {He}},
  \bibinfo {author} {\bibfnamefont {M.}~\bibnamefont {Le~Tacon}}, \bibinfo
  {author} {\bibfnamefont {I.~S.}\ \bibnamefont {Elfimov}}, \bibinfo {author}
  {\bibfnamefont {J.~E.}\ \bibnamefont {Hoffman}}, \bibinfo {author}
  {\bibfnamefont {G.~A.}\ \bibnamefont {Sawatzky}}, \bibinfo {author}
  {\bibfnamefont {B.}~\bibnamefont {Keimer}}, \ and\ \bibinfo {author}
  {\bibfnamefont {A.}~\bibnamefont {Damascelli}},\ }\href {\doibase
  10.1126/science.1242996} {\bibfield  {journal} {\bibinfo  {journal}
  {Science}\ }\textbf {\bibinfo {volume} {343}},\ \bibinfo {pages} {390}
  (\bibinfo {year} {2014})}\BibitemShut {NoStop}%
\bibitem [{\citenamefont {da~Silva~Neto}\ \emph {et~al.}(2015)\citenamefont
  {da~Silva~Neto}, \citenamefont {Comin}, \citenamefont {He}, \citenamefont
  {Sutarto}, \citenamefont {Jiang}, \citenamefont {Greene}, \citenamefont
  {Sawatzky},\ and\ \citenamefont {Damascelli}}]{daSilvaNeto282}%
  \BibitemOpen
  \bibfield  {author} {\bibinfo {author} {\bibfnamefont {E.~H.}\ \bibnamefont
  {da~Silva~Neto}}, \bibinfo {author} {\bibfnamefont {R.}~\bibnamefont
  {Comin}}, \bibinfo {author} {\bibfnamefont {F.}~\bibnamefont {He}}, \bibinfo
  {author} {\bibfnamefont {R.}~\bibnamefont {Sutarto}}, \bibinfo {author}
  {\bibfnamefont {Y.}~\bibnamefont {Jiang}}, \bibinfo {author} {\bibfnamefont
  {R.~L.}\ \bibnamefont {Greene}}, \bibinfo {author} {\bibfnamefont {G.~A.}\
  \bibnamefont {Sawatzky}}, \ and\ \bibinfo {author} {\bibfnamefont
  {A.}~\bibnamefont {Damascelli}},\ }\href {\doibase 10.1126/science.1256441}
  {\bibfield  {journal} {\bibinfo  {journal} {Science}\ }\textbf {\bibinfo
  {volume} {347}},\ \bibinfo {pages} {282} (\bibinfo {year}
  {2015})}\BibitemShut {NoStop}%
\bibitem [{\citenamefont {Wu}\ \emph {et~al.}(2012)\citenamefont {Wu},
  \citenamefont {Buchholz}, \citenamefont {Trabant}, \citenamefont {Chang},
  \citenamefont {Komarek}, \citenamefont {Heigl}, \citenamefont {Zimmermann},
  \citenamefont {Cwik}, \citenamefont {Nakamura}, \citenamefont {Braden},\ and\
  \citenamefont {Sch{\"u}{\ss}ler-Langeheine}}]{Wu2012}%
  \BibitemOpen
  \bibfield  {author} {\bibinfo {author} {\bibfnamefont {H.-H.}\ \bibnamefont
  {Wu}}, \bibinfo {author} {\bibfnamefont {M.}~\bibnamefont {Buchholz}},
  \bibinfo {author} {\bibfnamefont {C.}~\bibnamefont {Trabant}}, \bibinfo
  {author} {\bibfnamefont {C.~F.}\ \bibnamefont {Chang}}, \bibinfo {author}
  {\bibfnamefont {A.~C.}\ \bibnamefont {Komarek}}, \bibinfo {author}
  {\bibfnamefont {F.}~\bibnamefont {Heigl}}, \bibinfo {author} {\bibfnamefont
  {M.~v.}\ \bibnamefont {Zimmermann}}, \bibinfo {author} {\bibfnamefont
  {M.}~\bibnamefont {Cwik}}, \bibinfo {author} {\bibfnamefont {F.}~\bibnamefont
  {Nakamura}}, \bibinfo {author} {\bibfnamefont {M.}~\bibnamefont {Braden}}, \
  and\ \bibinfo {author} {\bibfnamefont {C.}~\bibnamefont
  {Sch{\"u}{\ss}ler-Langeheine}},\ }\href
  {http://dx.doi.org/10.1038/ncomms2019} {\bibfield  {journal} {\bibinfo
  {journal} {Nat. Commun.}\ }\textbf {\bibinfo {volume} {3}},\ \bibinfo {pages}
  {1023} (\bibinfo {year} {2012})}\BibitemShut {NoStop}%
\bibitem [{\citenamefont {Abbamonte}\ \emph {et~al.}(2005)\citenamefont
  {Abbamonte}, \citenamefont {Rusydi}, \citenamefont {Smadici}, \citenamefont
  {Gu}, \citenamefont {Sawatzky},\ and\ \citenamefont {Feng}}]{Abbamonte2005}%
  \BibitemOpen
  \bibfield  {author} {\bibinfo {author} {\bibfnamefont {P.}~\bibnamefont
  {Abbamonte}}, \bibinfo {author} {\bibfnamefont {A.}~\bibnamefont {Rusydi}},
  \bibinfo {author} {\bibfnamefont {S.}~\bibnamefont {Smadici}}, \bibinfo
  {author} {\bibfnamefont {G.~D.}\ \bibnamefont {Gu}}, \bibinfo {author}
  {\bibfnamefont {G.~A.}\ \bibnamefont {Sawatzky}}, \ and\ \bibinfo {author}
  {\bibfnamefont {D.~L.}\ \bibnamefont {Feng}},\ }\href {\doibase
  10.1038/nphys178} {\bibfield  {journal} {\bibinfo  {journal} {Nat. Phys.}\
  }\textbf {\bibinfo {volume} {1}},\ \bibinfo {pages} {155} (\bibinfo {year}
  {2005})}\BibitemShut {NoStop}%
\bibitem [{\citenamefont {Zaanen}\ and\ \citenamefont
  {Gunnarsson}(1989)}]{PhysRevB.40.7391}%
  \BibitemOpen
  \bibfield  {author} {\bibinfo {author} {\bibfnamefont {J.}~\bibnamefont
  {Zaanen}}\ and\ \bibinfo {author} {\bibfnamefont {O.}~\bibnamefont
  {Gunnarsson}},\ }\href {\doibase 10.1103/PhysRevB.40.7391} {\bibfield
  {journal} {\bibinfo  {journal} {Phys. Rev. B}\ }\textbf {\bibinfo {volume}
  {40}},\ \bibinfo {pages} {7391} (\bibinfo {year} {1989})}\BibitemShut
  {NoStop}%
\bibitem [{\citenamefont {Poilblanc}\ and\ \citenamefont
  {Rice}(1989)}]{PhysRevB.39.9749}%
  \BibitemOpen
  \bibfield  {author} {\bibinfo {author} {\bibfnamefont {D.}~\bibnamefont
  {Poilblanc}}\ and\ \bibinfo {author} {\bibfnamefont {T.~M.}\ \bibnamefont
  {Rice}},\ }\href {\doibase 10.1103/PhysRevB.39.9749} {\bibfield  {journal}
  {\bibinfo  {journal} {Phys. Rev. B}\ }\textbf {\bibinfo {volume} {39}},\
  \bibinfo {pages} {9749} (\bibinfo {year} {1989})}\BibitemShut {NoStop}%
\bibitem [{\citenamefont {Machida}(1989)}]{MACHIDA1989192}%
  \BibitemOpen
  \bibfield  {author} {\bibinfo {author} {\bibfnamefont {K.}~\bibnamefont
  {Machida}},\ }\href {\doibase http://dx.doi.org/10.1016/0921-4534(89)90316-X}
  {\bibfield  {journal} {\bibinfo  {journal} {Physica C}\ }\textbf {\bibinfo
  {volume} {158}},\ \bibinfo {pages} {192 } (\bibinfo {year}
  {1989})}\BibitemShut {NoStop}%
\bibitem [{\citenamefont {Tranquada}\ \emph {et~al.}(1995)\citenamefont
  {Tranquada}, \citenamefont {Sternlieb}, \citenamefont {Axe}, \citenamefont
  {Nakamura},\ and\ \citenamefont {Uchida}}]{Tranquada1995}%
  \BibitemOpen
  \bibfield  {author} {\bibinfo {author} {\bibfnamefont {J.~M.}\ \bibnamefont
  {Tranquada}}, \bibinfo {author} {\bibfnamefont {B.~J.}\ \bibnamefont
  {Sternlieb}}, \bibinfo {author} {\bibfnamefont {J.~D.}\ \bibnamefont {Axe}},
  \bibinfo {author} {\bibfnamefont {Y.}~\bibnamefont {Nakamura}}, \ and\
  \bibinfo {author} {\bibfnamefont {S.}~\bibnamefont {Uchida}},\ }\href
  {\doibase 10.1038/375561a0} {\bibfield  {journal} {\bibinfo  {journal}
  {Nature}\ }\textbf {\bibinfo {volume} {375}},\ \bibinfo {pages} {561}
  (\bibinfo {year} {1995})}\BibitemShut {NoStop}%
\bibitem [{\citenamefont {Bozin}\ \emph {et~al.}(2015)\citenamefont {Bozin},
  \citenamefont {Zhong}, \citenamefont {Knox}, \citenamefont {Gu},
  \citenamefont {Hill}, \citenamefont {Tranquada},\ and\ \citenamefont
  {Billinge}}]{PhysRevB.91.054521}%
  \BibitemOpen
  \bibfield  {author} {\bibinfo {author} {\bibfnamefont {E.~S.}\ \bibnamefont
  {Bozin}}, \bibinfo {author} {\bibfnamefont {R.}~\bibnamefont {Zhong}},
  \bibinfo {author} {\bibfnamefont {K.~R.}\ \bibnamefont {Knox}}, \bibinfo
  {author} {\bibfnamefont {G.}~\bibnamefont {Gu}}, \bibinfo {author}
  {\bibfnamefont {J.~P.}\ \bibnamefont {Hill}}, \bibinfo {author}
  {\bibfnamefont {J.~M.}\ \bibnamefont {Tranquada}}, \ and\ \bibinfo {author}
  {\bibfnamefont {S.~J.~L.}\ \bibnamefont {Billinge}},\ }\href {\doibase
  10.1103/PhysRevB.91.054521} {\bibfield  {journal} {\bibinfo  {journal} {Phys.
  Rev. B}\ }\textbf {\bibinfo {volume} {91}},\ \bibinfo {pages} {054521}
  (\bibinfo {year} {2015})}\BibitemShut {NoStop}%
\bibitem [{\citenamefont {Udby}\ \emph {et~al.}(2013)\citenamefont {Udby},
  \citenamefont {Larsen}, \citenamefont {Christensen}, \citenamefont {Boehm},
  \citenamefont {Niedermayer}, \citenamefont {Mohottala}, \citenamefont
  {Jensen}, \citenamefont {Toft-Petersen}, \citenamefont {Chou}, \citenamefont
  {Andersen}, \citenamefont {Lefmann},\ and\ \citenamefont
  {Wells}}]{PhysRevLett.111.227001}%
  \BibitemOpen
  \bibfield  {author} {\bibinfo {author} {\bibfnamefont {L.}~\bibnamefont
  {Udby}}, \bibinfo {author} {\bibfnamefont {J.}~\bibnamefont {Larsen}},
  \bibinfo {author} {\bibfnamefont {N.~B.}\ \bibnamefont {Christensen}},
  \bibinfo {author} {\bibfnamefont {M.}~\bibnamefont {Boehm}}, \bibinfo
  {author} {\bibfnamefont {C.}~\bibnamefont {Niedermayer}}, \bibinfo {author}
  {\bibfnamefont {H.~E.}\ \bibnamefont {Mohottala}}, \bibinfo {author}
  {\bibfnamefont {T.~B.~S.}\ \bibnamefont {Jensen}}, \bibinfo {author}
  {\bibfnamefont {R.}~\bibnamefont {Toft-Petersen}}, \bibinfo {author}
  {\bibfnamefont {F.~C.}\ \bibnamefont {Chou}}, \bibinfo {author}
  {\bibfnamefont {N.~H.}\ \bibnamefont {Andersen}}, \bibinfo {author}
  {\bibfnamefont {K.}~\bibnamefont {Lefmann}}, \ and\ \bibinfo {author}
  {\bibfnamefont {B.~O.}\ \bibnamefont {Wells}},\ }\href {\doibase
  10.1103/PhysRevLett.111.227001} {\bibfield  {journal} {\bibinfo  {journal}
  {Phys. Rev. Lett.}\ }\textbf {\bibinfo {volume} {111}},\ \bibinfo {pages}
  {227001} (\bibinfo {year} {2013})}\BibitemShut {NoStop}%
\bibitem [{\citenamefont {Mohottala}\ \emph {et~al.}(2006)\citenamefont
  {Mohottala}, \citenamefont {Wells}, \citenamefont {Budnick}, \citenamefont
  {Hines}, \citenamefont {Niedermayer}, \citenamefont {Udby}, \citenamefont
  {Bernhard}, \citenamefont {Moodenbaugh},\ and\ \citenamefont
  {Chou}}]{Mohottala2006}%
  \BibitemOpen
  \bibfield  {author} {\bibinfo {author} {\bibfnamefont {H.~E.}\ \bibnamefont
  {Mohottala}}, \bibinfo {author} {\bibfnamefont {B.~O.}\ \bibnamefont
  {Wells}}, \bibinfo {author} {\bibfnamefont {J.~I.}\ \bibnamefont {Budnick}},
  \bibinfo {author} {\bibfnamefont {W.~A.}\ \bibnamefont {Hines}}, \bibinfo
  {author} {\bibfnamefont {C.}~\bibnamefont {Niedermayer}}, \bibinfo {author}
  {\bibfnamefont {L.}~\bibnamefont {Udby}}, \bibinfo {author} {\bibfnamefont
  {C.}~\bibnamefont {Bernhard}}, \bibinfo {author} {\bibfnamefont {A.~R.}\
  \bibnamefont {Moodenbaugh}}, \ and\ \bibinfo {author} {\bibfnamefont {F.-C.}\
  \bibnamefont {Chou}},\ }\href {\doibase 10.1038/nmat1633} {\bibfield
  {journal} {\bibinfo  {journal} {Nat. Mater.}\ }\textbf {\bibinfo {volume}
  {5}},\ \bibinfo {pages} {377} (\bibinfo {year} {2006})}\BibitemShut {NoStop}%
\bibitem [{\citenamefont {Chen}\ \emph {et~al.}(1991)\citenamefont {Chen},
  \citenamefont {Sette}, \citenamefont {Ma}, \citenamefont {Hybertsen},
  \citenamefont {Stechel}, \citenamefont {Foulkes}, \citenamefont {Schulter},
  \citenamefont {Cheong}, \citenamefont {Cooper}, \citenamefont {Rupp},
  \citenamefont {Batlogg}, \citenamefont {Soo}, \citenamefont {Ming},
  \citenamefont {Krol},\ and\ \citenamefont {Kao}}]{ctchen}%
  \BibitemOpen
  \bibfield  {author} {\bibinfo {author} {\bibfnamefont {C. T.}~\bibnamefont
  {Chen}}, \bibinfo {author} {\bibfnamefont {F.}~\bibnamefont {Sette}},
  \bibinfo {author} {\bibfnamefont {Y.}~\bibnamefont {Ma}}, \bibinfo {author}
  {\bibfnamefont {M.}~\bibnamefont {Hybertsen}}, \bibinfo {author}
  {\bibfnamefont {E.}~\bibnamefont {Stechel}}, \bibinfo {author} {\bibfnamefont
  {W.}~\bibnamefont {Foulkes}}, \bibinfo {author} {\bibfnamefont
  {M.}~\bibnamefont {Schulter}}, \bibinfo {author} {\bibfnamefont
  {S.}~\bibnamefont {Cheong}}, \bibinfo {author} {\bibfnamefont
  {A.}~\bibnamefont {Cooper}}, \bibinfo {author} {\bibfnamefont
  {L.}~\bibnamefont {Rupp}}, \bibinfo {author} {\bibfnamefont {B.}~\bibnamefont
  {Batlogg}}, \bibinfo {author} {\bibfnamefont {Y.}~\bibnamefont {Soo}},
  \bibinfo {author} {\bibfnamefont {Z.}~\bibnamefont {Ming}}, \bibinfo {author}
  {\bibfnamefont {A.}~\bibnamefont {Krol}}, \ and\ \bibinfo {author}
  {\bibfnamefont {Y.}~\bibnamefont {Kao}},\ }\href {\doibase
  10.1103/PhysRevLett.66.104} {\bibfield  {journal} {\bibinfo  {journal} {Phys.
  Rev. Lett.}\ }\textbf {\bibinfo {volume} {66}},\ \bibinfo {pages} {104}
  (\bibinfo {year} {1991})}\BibitemShut {NoStop}%
\bibitem [{sup()}]{supp}%
  \BibitemOpen
  \href@noop {} {}\bibinfo {note} {See Supplemental Material for supportive
  details.}\BibitemShut {Stop}%
\bibitem [{\citenamefont {Achkar}\ \emph {et~al.}(2013)\citenamefont {Achkar},
  \citenamefont {He}, \citenamefont {Sutarto}, \citenamefont {Geck},
  \citenamefont {Zhang}, \citenamefont {Kim},\ and\ \citenamefont
  {Hawthorn}}]{PhysRevLett.110.017001}%
  \BibitemOpen
  \bibfield  {author} {\bibinfo {author} {\bibfnamefont {A.~J.}\ \bibnamefont
  {Achkar}}, \bibinfo {author} {\bibfnamefont {F.}~\bibnamefont {He}}, \bibinfo
  {author} {\bibfnamefont {R.}~\bibnamefont {Sutarto}}, \bibinfo {author}
  {\bibfnamefont {J.}~\bibnamefont {Geck}}, \bibinfo {author} {\bibfnamefont
  {H.}~\bibnamefont {Zhang}}, \bibinfo {author} {\bibfnamefont {Y.-J.}\
  \bibnamefont {Kim}}, \ and\ \bibinfo {author} {\bibfnamefont {D.~G.}\
  \bibnamefont {Hawthorn}},\ }\href {\doibase 10.1103/PhysRevLett.110.017001}
  {\bibfield  {journal} {\bibinfo  {journal} {Phys. Rev. Lett.}\ }\textbf
  {\bibinfo {volume} {110}},\ \bibinfo {pages} {017001} (\bibinfo {year}
  {2013})}\BibitemShut {NoStop}%
\bibitem [{\citenamefont {H\"ucker}\ \emph {et~al.}(2011)\citenamefont
  {H\"ucker}, \citenamefont {v.~Zimmermann}, \citenamefont {Gu}, \citenamefont
  {Xu}, \citenamefont {Wen}, \citenamefont {Xu}, \citenamefont {Kang},
  \citenamefont {Zheludev},\ and\ \citenamefont
  {Tranquada}}]{PhysRevB.83.104506}%
  \BibitemOpen
  \bibfield  {author} {\bibinfo {author} {\bibfnamefont {M.}~\bibnamefont
  {H\"ucker}}, \bibinfo {author} {\bibfnamefont {M.}~\bibnamefont
  {v.~Zimmermann}}, \bibinfo {author} {\bibfnamefont {G.~D.}\ \bibnamefont
  {Gu}}, \bibinfo {author} {\bibfnamefont {Z.~J.}\ \bibnamefont {Xu}}, \bibinfo
  {author} {\bibfnamefont {J.~S.}\ \bibnamefont {Wen}}, \bibinfo {author}
  {\bibfnamefont {G.}~\bibnamefont {Xu}}, \bibinfo {author} {\bibfnamefont
  {H.~J.}\ \bibnamefont {Kang}}, \bibinfo {author} {\bibfnamefont
  {A.}~\bibnamefont {Zheludev}}, \ and\ \bibinfo {author} {\bibfnamefont
  {J.~M.}\ \bibnamefont {Tranquada}},\ }\href {\doibase
  10.1103/PhysRevB.83.104506} {\bibfield  {journal} {\bibinfo  {journal} {Phys.
  Rev. B}\ }\textbf {\bibinfo {volume} {83}},\ \bibinfo {pages} {104506}
  (\bibinfo {year} {2011})}\BibitemShut {NoStop}%
\bibitem [{\citenamefont {Fink}\ \emph {et~al.}(2011)\citenamefont {Fink},
  \citenamefont {Soltwisch}, \citenamefont {Geck}, \citenamefont {Schierle},
  \citenamefont {Weschke},\ and\ \citenamefont
  {B\"uchner}}]{PhysRevB.83.092503}%
  \BibitemOpen
  \bibfield  {author} {\bibinfo {author} {\bibfnamefont {J.}~\bibnamefont
  {Fink}}, \bibinfo {author} {\bibfnamefont {V.}~\bibnamefont {Soltwisch}},
  \bibinfo {author} {\bibfnamefont {J.}~\bibnamefont {Geck}}, \bibinfo {author}
  {\bibfnamefont {E.}~\bibnamefont {Schierle}}, \bibinfo {author}
  {\bibfnamefont {E.}~\bibnamefont {Weschke}}, \ and\ \bibinfo {author}
  {\bibfnamefont {B.}~\bibnamefont {B\"uchner}},\ }\href {\doibase
  10.1103/PhysRevB.83.092503} {\bibfield  {journal} {\bibinfo  {journal} {Phys.
  Rev. B}\ }\textbf {\bibinfo {volume} {83}},\ \bibinfo {pages} {092503}
  (\bibinfo {year} {2011})}\BibitemShut {NoStop}%
\bibitem [{\citenamefont {Christensen}\ \emph {et~al.}(2007)\citenamefont
  {Christensen}, \citenamefont {R\o{}nnow}, \citenamefont {Mesot},
  \citenamefont {Ewings}, \citenamefont {Momono}, \citenamefont {Oda},
  \citenamefont {Ido}, \citenamefont {Enderle}, \citenamefont {McMorrow},\ and\
  \citenamefont {Boothroyd}}]{PhysRevLett.98.197003}%
  \BibitemOpen
  \bibfield  {author} {\bibinfo {author} {\bibfnamefont {N.~B.}\ \bibnamefont
  {Christensen}}, \bibinfo {author} {\bibfnamefont {H.~M.}\ \bibnamefont
  {R\o{}nnow}}, \bibinfo {author} {\bibfnamefont {J.}~\bibnamefont {Mesot}},
  \bibinfo {author} {\bibfnamefont {R.~A.}\ \bibnamefont {Ewings}}, \bibinfo
  {author} {\bibfnamefont {N.}~\bibnamefont {Momono}}, \bibinfo {author}
  {\bibfnamefont {M.}~\bibnamefont {Oda}}, \bibinfo {author} {\bibfnamefont
  {M.}~\bibnamefont {Ido}}, \bibinfo {author} {\bibfnamefont {M.}~\bibnamefont
  {Enderle}}, \bibinfo {author} {\bibfnamefont {D.~F.}\ \bibnamefont
  {McMorrow}}, \ and\ \bibinfo {author} {\bibfnamefont {A.~T.}\ \bibnamefont
  {Boothroyd}},\ }\href {\doibase 10.1103/PhysRevLett.98.197003} {\bibfield
  {journal} {\bibinfo  {journal} {Phys. Rev. Lett.}\ }\textbf {\bibinfo
  {volume} {98}},\ \bibinfo {pages} {197003} (\bibinfo {year}
  {2007})}\BibitemShut {NoStop}%
\bibitem [{\citenamefont {Zimmermann}\ \emph {et~al.}(1998)\citenamefont
  {Zimmermann}, \citenamefont {Vigliante}, \citenamefont {Niemöller},
  \citenamefont {Ichikawa}, \citenamefont {Frello}, \citenamefont {Madsen},
  \citenamefont {Wochner}, \citenamefont {Uchida}, \citenamefont {Andersen},
  \citenamefont {Tranquada}, \citenamefont {Gibbs},\ and\ \citenamefont
  {Schneider}}]{zimmerman}%
  \BibitemOpen
  \bibfield  {author} {\bibinfo {author} {\bibfnamefont {M.}~\bibnamefont
  {Zimmermann}}, \bibinfo {author} {\bibfnamefont {A.}~\bibnamefont
  {Vigliante}}, \bibinfo {author} {\bibfnamefont {T.}~\bibnamefont
  {Niemöller}}, \bibinfo {author} {\bibfnamefont {N.}~\bibnamefont
  {Ichikawa}}, \bibinfo {author} {\bibfnamefont {T.}~\bibnamefont {Frello}},
  \bibinfo {author} {\bibfnamefont {J.}~\bibnamefont {Madsen}}, \bibinfo
  {author} {\bibfnamefont {P.}~\bibnamefont {Wochner}}, \bibinfo {author}
  {\bibfnamefont {S.}~\bibnamefont {Uchida}}, \bibinfo {author} {\bibfnamefont
  {N.}~\bibnamefont {Andersen}}, \bibinfo {author} {\bibfnamefont
  {J.}~\bibnamefont {Tranquada}}, \bibinfo {author} {\bibfnamefont
  {D.}~\bibnamefont {Gibbs}}, \ and\ \bibinfo {author} {\bibfnamefont
  {J.}~\bibnamefont {Schneider}},\ }\href@noop {} {\bibfield  {journal}
  {\bibinfo  {journal} {Europhysics Letters}\ }\textbf {\bibinfo {volume}
  {41}},\ \bibinfo {pages} {629} (\bibinfo {year} {1998})}\BibitemShut
  {NoStop}%
\bibitem [{\citenamefont {Ray}\ \emph {et~al.}(2017)\citenamefont {Ray},
  \citenamefont {Andersen}, \citenamefont {Jensen}, \citenamefont {Mohottala},
  \citenamefont {Niedermayer}, \citenamefont {Lefmann}, \citenamefont {Wells},
  \citenamefont {v.~Zimmermann},\ and\ \citenamefont {Udby}}]{1707.08871}%
  \BibitemOpen
  \bibfield  {author} {\bibinfo {author} {\bibfnamefont {P.~J.}\ \bibnamefont
  {Ray}}, \bibinfo {author} {\bibfnamefont {N.~H.}\ \bibnamefont {Andersen}},
  \bibinfo {author} {\bibfnamefont {T.~B.~S.}\ \bibnamefont {Jensen}}, \bibinfo
  {author} {\bibfnamefont {H.~E.}\ \bibnamefont {Mohottala}}, \bibinfo {author}
  {\bibfnamefont {C.}~\bibnamefont {Niedermayer}}, \bibinfo {author}
  {\bibfnamefont {K.}~\bibnamefont {Lefmann}}, \bibinfo {author} {\bibfnamefont
  {B.~O.}\ \bibnamefont {Wells}}, \bibinfo {author} {\bibfnamefont
  {M.}~\bibnamefont {v.~Zimmermann}}, \ and\ \bibinfo {author} {\bibfnamefont
  {L.}~\bibnamefont {Udby}},\ }\href@noop {} (\bibinfo {year} {2017}),\ \Eprint
  {http://arxiv.org/abs/arXiv:1707.08871} {arXiv:1707.08871} \BibitemShut
  {NoStop}%
\bibitem [{\citenamefont {Wells}\ \emph {et~al.}(1996)\citenamefont {Wells},
  \citenamefont {Birgeneau}, \citenamefont {Chou}, \citenamefont {Endoh},
  \citenamefont {Johnston}, \citenamefont {Kastner}, \citenamefont {Lee},
  \citenamefont {Shirane}, \citenamefont {Tranquada},\ and\ \citenamefont
  {Yamada}}]{Wells1996}%
  \BibitemOpen
  \bibfield  {author} {\bibinfo {author} {\bibfnamefont {B.~O.}\ \bibnamefont
  {Wells}}, \bibinfo {author} {\bibfnamefont {R.~J.}\ \bibnamefont
  {Birgeneau}}, \bibinfo {author} {\bibfnamefont {F.~C.}\ \bibnamefont {Chou}},
  \bibinfo {author} {\bibfnamefont {Y.}~\bibnamefont {Endoh}}, \bibinfo
  {author} {\bibfnamefont {D.~C.}\ \bibnamefont {Johnston}}, \bibinfo {author}
  {\bibfnamefont {M.~A.}\ \bibnamefont {Kastner}}, \bibinfo {author}
  {\bibfnamefont {Y.~S.}\ \bibnamefont {Lee}}, \bibinfo {author} {\bibfnamefont
  {G.}~\bibnamefont {Shirane}}, \bibinfo {author} {\bibfnamefont {J.~M.}\
  \bibnamefont {Tranquada}}, \ and\ \bibinfo {author} {\bibfnamefont
  {K.}~\bibnamefont {Yamada}},\ }\href {\doibase 10.1007/s002570050158}
  {\bibfield  {journal} {\bibinfo  {journal} {Z. Phys. B}\ }\textbf {\bibinfo
  {volume} {100}},\ \bibinfo {pages} {535} (\bibinfo {year}
  {1996})}\BibitemShut {NoStop}%
\bibitem [{\citenamefont {Yamada}\ \emph {et~al.}(1998)\citenamefont {Yamada},
  \citenamefont {Lee}, \citenamefont {Kurahashi}, \citenamefont {Wada},
  \citenamefont {Wakimoto}, \citenamefont {Ueki}, \citenamefont {Kimura},
  \citenamefont {Endoh}, \citenamefont {Hosoya}, \citenamefont {Shirane},
  \citenamefont {Birgeneau}, \citenamefont {Greven}, \citenamefont {Kastner},\
  and\ \citenamefont {Kim}}]{PhysRevB.57.6165}%
  \BibitemOpen
  \bibfield  {author} {\bibinfo {author} {\bibfnamefont {K.}~\bibnamefont
  {Yamada}}, \bibinfo {author} {\bibfnamefont {C.~H.}\ \bibnamefont {Lee}},
  \bibinfo {author} {\bibfnamefont {K.}~\bibnamefont {Kurahashi}}, \bibinfo
  {author} {\bibfnamefont {J.}~\bibnamefont {Wada}}, \bibinfo {author}
  {\bibfnamefont {S.}~\bibnamefont {Wakimoto}}, \bibinfo {author}
  {\bibfnamefont {S.}~\bibnamefont {Ueki}}, \bibinfo {author} {\bibfnamefont
  {H.}~\bibnamefont {Kimura}}, \bibinfo {author} {\bibfnamefont
  {Y.}~\bibnamefont {Endoh}}, \bibinfo {author} {\bibfnamefont
  {S.}~\bibnamefont {Hosoya}}, \bibinfo {author} {\bibfnamefont
  {G.}~\bibnamefont {Shirane}}, \bibinfo {author} {\bibfnamefont {R.~J.}\
  \bibnamefont {Birgeneau}}, \bibinfo {author} {\bibfnamefont {M.}~\bibnamefont
  {Greven}}, \bibinfo {author} {\bibfnamefont {M.~A.}\ \bibnamefont {Kastner}},
  \ and\ \bibinfo {author} {\bibfnamefont {Y.~J.}\ \bibnamefont {Kim}},\ }\href
  {\doibase 10.1103/PhysRevB.57.6165} {\bibfield  {journal} {\bibinfo
  {journal} {Phys. Rev. B}\ }\textbf {\bibinfo {volume} {57}},\ \bibinfo
  {pages} {6165} (\bibinfo {year} {1998})}\BibitemShut {NoStop}%
\bibitem [{\citenamefont {Tranquada}\ \emph {et~al.}(2008)\citenamefont
  {Tranquada}, \citenamefont {Gu}, \citenamefont {H\"ucker}, \citenamefont
  {Jie}, \citenamefont {Kang}, \citenamefont {Klingeler}, \citenamefont {Li},
  \citenamefont {Tristan}, \citenamefont {Wen}, \citenamefont {Xu},
  \citenamefont {Xu}, \citenamefont {Zhou},\ and\ \citenamefont
  {v.~Zimmermann}}]{PhysRevB.78.174529}%
  \BibitemOpen
  \bibfield  {author} {\bibinfo {author} {\bibfnamefont {J.~M.}\ \bibnamefont
  {Tranquada}}, \bibinfo {author} {\bibfnamefont {G.~D.}\ \bibnamefont {Gu}},
  \bibinfo {author} {\bibfnamefont {M.}~\bibnamefont {H\"ucker}}, \bibinfo
  {author} {\bibfnamefont {Q.}~\bibnamefont {Jie}}, \bibinfo {author}
  {\bibfnamefont {H.-J.}\ \bibnamefont {Kang}}, \bibinfo {author}
  {\bibfnamefont {R.}~\bibnamefont {Klingeler}}, \bibinfo {author}
  {\bibfnamefont {Q.}~\bibnamefont {Li}}, \bibinfo {author} {\bibfnamefont
  {N.}~\bibnamefont {Tristan}}, \bibinfo {author} {\bibfnamefont {J.~S.}\
  \bibnamefont {Wen}}, \bibinfo {author} {\bibfnamefont {G.~Y.}\ \bibnamefont
  {Xu}}, \bibinfo {author} {\bibfnamefont {Z.~J.}\ \bibnamefont {Xu}}, \bibinfo
  {author} {\bibfnamefont {J.}~\bibnamefont {Zhou}}, \ and\ \bibinfo {author}
  {\bibfnamefont {M.}~\bibnamefont {v.~Zimmermann}},\ }\href {\doibase
  10.1103/PhysRevB.78.174529} {\bibfield  {journal} {\bibinfo  {journal} {Phys.
  Rev. B}\ }\textbf {\bibinfo {volume} {78}},\ \bibinfo {pages} {174529}
  (\bibinfo {year} {2008})}\BibitemShut {NoStop}%
\bibitem [{\citenamefont {Klauss}\ \emph {et~al.}(2000)\citenamefont {Klauss},
  \citenamefont {Wagener}, \citenamefont {Hillberg}, \citenamefont {Kopmann},
  \citenamefont {Walf}, \citenamefont {Litterst}, \citenamefont {H\"ucker},\
  and\ \citenamefont {B\"uchner}}]{PhysRevLett.85.4590}%
  \BibitemOpen
  \bibfield  {author} {\bibinfo {author} {\bibfnamefont {H.-H.}\ \bibnamefont
  {Klauss}}, \bibinfo {author} {\bibfnamefont {W.}~\bibnamefont {Wagener}},
  \bibinfo {author} {\bibfnamefont {M.}~\bibnamefont {Hillberg}}, \bibinfo
  {author} {\bibfnamefont {W.}~\bibnamefont {Kopmann}}, \bibinfo {author}
  {\bibfnamefont {H.}~\bibnamefont {Walf}}, \bibinfo {author} {\bibfnamefont
  {F.~J.}\ \bibnamefont {Litterst}}, \bibinfo {author} {\bibfnamefont
  {M.}~\bibnamefont {H\"ucker}}, \ and\ \bibinfo {author} {\bibfnamefont
  {B.}~\bibnamefont {B\"uchner}},\ }\href {\doibase
  10.1103/PhysRevLett.85.4590} {\bibfield  {journal} {\bibinfo  {journal}
  {Phys. Rev. Lett.}\ }\textbf {\bibinfo {volume} {85}},\ \bibinfo {pages}
  {4590} (\bibinfo {year} {2000})}\BibitemShut {NoStop}%
\bibitem [{\citenamefont {Christensen}\ \emph {et~al.}(2014)\citenamefont
  {Christensen}, \citenamefont {Chang}, \citenamefont {Larsen}, \citenamefont
  {Fujita}, \citenamefont {Oda}, \citenamefont {Ido}, \citenamefont {Momono},
  \citenamefont {Forgan}, \citenamefont {Holmes}, \citenamefont {Mesot},
  \citenamefont {Huecker},\ and\ \citenamefont {v.~Zimmermann}}]{1404.3192}%
  \BibitemOpen
  \bibfield  {author} {\bibinfo {author} {\bibfnamefont {N.~B.}\ \bibnamefont
  {Christensen}}, \bibinfo {author} {\bibfnamefont {J.}~\bibnamefont {Chang}},
  \bibinfo {author} {\bibfnamefont {J.}~\bibnamefont {Larsen}}, \bibinfo
  {author} {\bibfnamefont {M.}~\bibnamefont {Fujita}}, \bibinfo {author}
  {\bibfnamefont {M.}~\bibnamefont {Oda}}, \bibinfo {author} {\bibfnamefont
  {M.}~\bibnamefont {Ido}}, \bibinfo {author} {\bibfnamefont {N.}~\bibnamefont
  {Momono}}, \bibinfo {author} {\bibfnamefont {E.~M.}\ \bibnamefont {Forgan}},
  \bibinfo {author} {\bibfnamefont {A.~T.}\ \bibnamefont {Holmes}}, \bibinfo
  {author} {\bibfnamefont {J.}~\bibnamefont {Mesot}}, \bibinfo {author}
  {\bibfnamefont {M.}~\bibnamefont {Huecker}}, \ and\ \bibinfo {author}
  {\bibfnamefont {M.}~\bibnamefont {v.~Zimmermann}},\ }\href@noop {} {\
  (\bibinfo {year} {2014})},\ \Eprint {http://arxiv.org/abs/arXiv:1404.3192}
  {arXiv:1404.3192} \BibitemShut {NoStop}%
\bibitem [{\citenamefont {Croft}\ \emph {et~al.}(2014)\citenamefont {Croft},
  \citenamefont {Lester}, \citenamefont {Senn}, \citenamefont {Bombardi},\ and\
  \citenamefont {Hayden}}]{PhysRevB.89.224513}%
  \BibitemOpen
  \bibfield  {author} {\bibinfo {author} {\bibfnamefont {T.~P.}\ \bibnamefont
  {Croft}}, \bibinfo {author} {\bibfnamefont {C.}~\bibnamefont {Lester}},
  \bibinfo {author} {\bibfnamefont {M.~S.}\ \bibnamefont {Senn}}, \bibinfo
  {author} {\bibfnamefont {A.}~\bibnamefont {Bombardi}}, \ and\ \bibinfo
  {author} {\bibfnamefont {S.~M.}\ \bibnamefont {Hayden}},\ }\href {\doibase
  10.1103/PhysRevB.89.224513} {\bibfield  {journal} {\bibinfo  {journal} {Phys.
  Rev. B}\ }\textbf {\bibinfo {volume} {89}},\ \bibinfo {pages} {224513}
  (\bibinfo {year} {2014})}\BibitemShut {NoStop}%
\end{thebibliography}

\end{document}